\documentclass[12pt,preprint]{aastex}
\usepackage{epsfig}

\begin{document}

\title{The Expanded Very Large Array}

\author{Richard Perley, Peter Napier, Jim Jackson, Bryan Butler, Barry Clark, Robert 
        Hayward, Steven Durand, Mike Revnell, and Mark McKinnon}
\affil{National Radio Astronomy Observatory, Socorro, NM 87801 USA}
\author{Brent Carlson and David Fort}
\affil{Dominion Radio Astrophysical Observatory, Penticton, BC V2A 6K3 Canada}
\and
\author{Peter Dewdney}
\affil{University of Manchester, Manchester, UK M13 9PL}

\begin{abstract}

In almost 30 years of operation, the Very Large Array (VLA) has proved to be a 
remarkably flexible and productive radio telescope. However, the basic 
capabilities of the VLA have changed little since it was designed. A major 
expansion utilizing modern technology is currently underway to improve the 
capabilities of the VLA by at least an order of magnitude in both sensitivity 
and in frequency coverage. The primary elements of the Expanded Very Large 
Array (EVLA) project include new or upgraded receivers for continuous 
frequency coverage from 1 to 50 GHz, new local oscillator, intermediate 
frequency, and wide bandwidth data transmission systems to carry signals with 
16 GHz total bandwidth from each antenna, and a new digital correlator with 
the capability to process this bandwidth with an unprecedented number of 
frequency channels for an imaging array. Also included are a new monitor and 
control system and new software that will provide telescope ease of use. 
Scheduled for completion in 2012, the EVLA will provide the world research 
community with a flexible, powerful, general-purpose telescope to address 
current and future astronomical issues.

\end{abstract}


\section{Introduction}

The Very Large Array (VLA) is an imaging array located on the plains of San 
Augustin in west-central New Mexico. It consists of a total of 27 antennas, 
each 25 meters in diameter, with nine antennas distributed along each of 
three equilangular arms extending out to 21 km from the center. The array 
provides diffraction-limited images of astronomical objects in all Stokes 
parameters, with a maximum resolution at 1.4 GHz of 1.4 arcseconds, and at 
45 GHz of 0.05 arcseconds. Detailed descriptions of the VLA as originally 
constructed can be found in [1] and [2]. 

The VLA was designed and built in the 1970s, and utilized the best technology 
of that time. The telescope, upon completion in 1980, could observe in four 
frequency bands, utilizing state-of-the-art cryogenically cooled receivers, 
TE01 circular waveguide to transport the analog signals of 200 MHz bandwidth 
from the most distant antennas to the central location without amplification 
and, most notably, a digital correlator capable of producing up to 512 
spectral channels spanning 3 MHz for each of the 351 baselines, or 
full-polarization continuum correlations at two frequencies simultaneously 
with 50 MHz bandwidth. The VLA increased astronomical capabilities by one or 
more orders of magnitude over all preceding radio telescopes in sensitivity, 
resolution, frequency coverage, speed, flexibility, and imaging fidelity. 

The design of the VLA was heavily influenced by both the available technology 
of the 1970s, and the key scientific questions of that era, which included 
imaging the Doppler-shifted emission from neutral hydrogen from local galaxies 
and resolving the bright continuum emission of distant quasars, radio galaxies,
and supernova remnants. The VLA has been spectacularly successful both in 
addressing these questions, and in its application to a wide range of 
astrophysical problems unknown or unanticipated in the 1970s.  It has been so 
successful in this latter area because it was designed as a general purpose, 
reconfigurable array. However, the VLA has changed very little since 1980.  
Although most of the receiver bands have seen improvements in sensitivity, 
and four new receiving systems have been added, the array's original signal 
transmission system and correlator remain unchanged. In the approximately 
30 years since the VLA was designed, there have been enormous advances in 
technology, particularly in digital communication and signal processing 
capabilities. Over the same interval, the key scientific questions have also 
undergone great changes, requiring telescopes with ever greater emphasis on 
sensitivity, wider frequency coverage, faster surveying capabilities, more 
spectral capabilities, higher imaging fidelity, and faster response to 
transient emission. One can respond to these challenges in two ways: by 
designing wholly new instruments, or by utilizing modern technologies to 
upgrade, and expand, the world's preeminent existing telescopes. Taking the 
latter approach has resulted in the Expanded Very Large Array (EVLA) project.

The technical requirements for the EVLA are based on a comprehensive review 
of the potential science enabled by order of magnitude, or greater, 
improvements over existing VLA capabilities. There are four major science 
themes for the EVLA:

\begin{itemize}
\item{The Magnetic Universe: Measuring the strength and topology of magnetic 
      fields,}

\item{The Obscured Universe: Enabling unbiased surveys, and imaging of 
      dust-shrouded objects which are obscured at other wavelengths,}

\item{The Transient Universe: Enabling rapid response to, and imaging of, 
      rapidly evolving transient sources, and}

\item{The Evolving Universe: Tracking the formation and evolution of objects 
      in our universe, ranging from stars to spiral galaxies and galactic 
      nuclei.}
\end{itemize}

For all of these themes, it was readily demonstrated that order-of-magnitude 
improvements in VLA performance by implementation of modern technologies will 
result in spectacular new science by the world user community.  The EVLA is a 
comprehensive technical upgrade of the VLA.  It is a leveraged project: by 
utilizing relatively inexpensive modern digital electronics, and building on 
the existing infrastructure of the VLA, the EVLA will advance by one to four 
orders of magnitude all the imaging capabilities of the VLA at modest cost and 
in a time short compared to that needed to design and build a new facility.
The EVLA will provide astronomers a modern, general-purpose, radio telescope 
capable of addressing the key scientific issues of the day, and the 
yet-unforeseen issues of the future.

The EVLA project was started in 2001. The new correlator will be installed in 
2009, the conversion of the VLA antennas to the EVLA design will be complete 
in 2010, and all receivers will be installed by 2012. The entire project will 
be complete in 2012. The project is funded by the National Science Foundation 
of the United States of America, the National Research Council of Canada, and 
the Consejo Nacional de Ciencia y Tecnologia of Mexico.

An overview of the system is described in Section~\ref{sec:overview}. A more 
detailed description of the instrument's major subsystems, including feed horns 
and receivers, local oscillator, intermediate frequency, and data transmission 
systems, and software, is given in Section~\ref{sec:subsystems}. The paper is 
summarized in Section~\ref{sec:summary}. The descriptions of acronyms used 
throughout the paper are listed in the Glossary.

\begin{figure}
\begin{center}
\epsfig{file=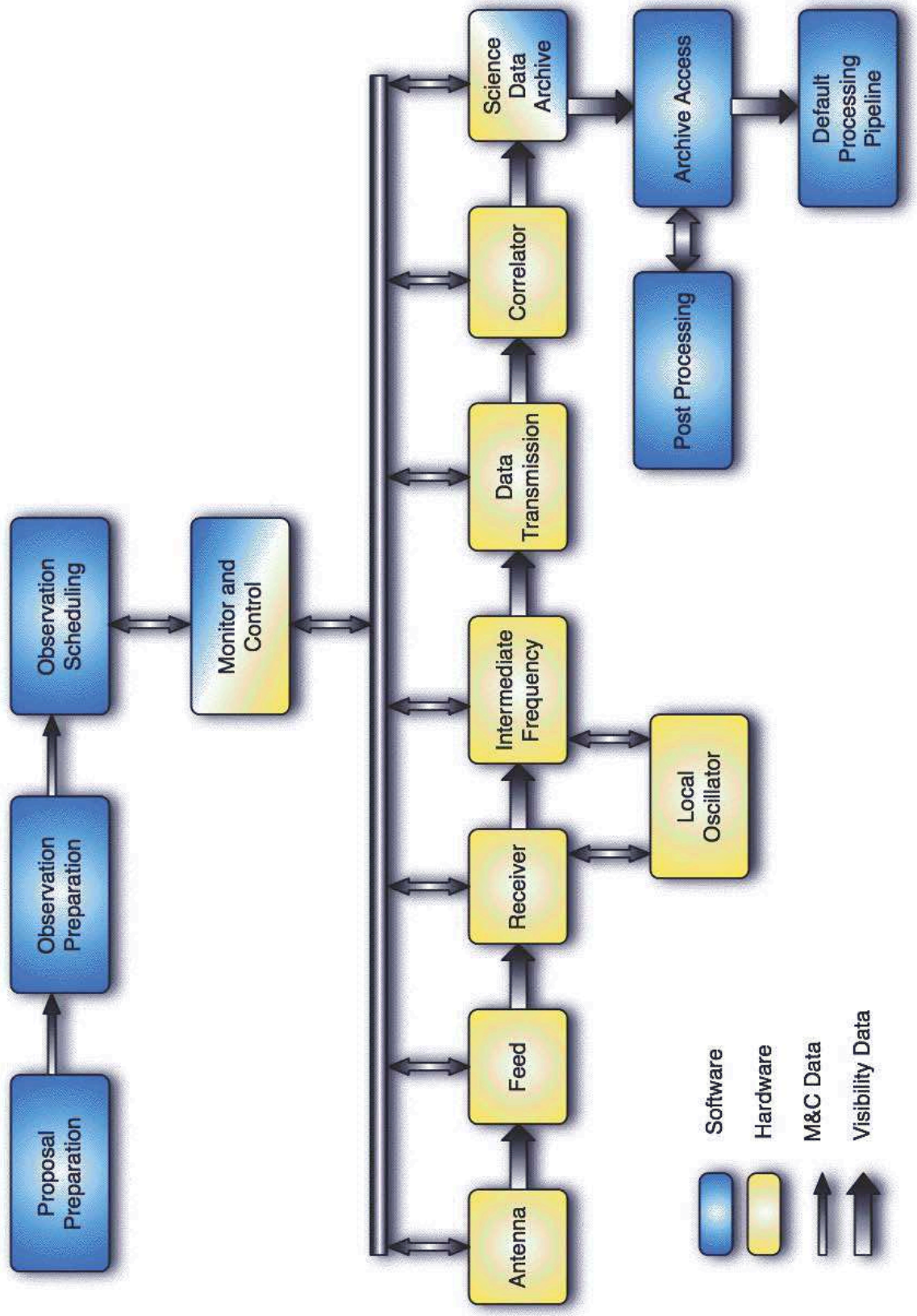, height=6in, angle=90}
\end{center}
\caption{Simplified block diagram for the EVLA}
\label{fig:blockdiagram}
\end{figure}

\section{System Overview}
\label{sec:overview}

The primary technical requirements for the EVLA that were determined from 
the instrument's scientific goals are:

\begin{itemize}

\item{New or upgraded receivers at the Cassegrain focus of the antennas 
      providing continuous frequency coverage from 1 to 50 GHz in 8 frequency 
      bands (see Table 1).}

\item{A new, wide bandwidth, fiber optic data transmission system, including 
      associated local oscillator (LO) and intermediate frequency (IF) 
      electronics, to carry signals with 16 GHz total bandwidth from each 
      antenna to the correlator.}

\item{New electronics to process 8 signal channels of up to 2 GHz bandwidth 
      each.}

\item{A new, wide bandwidth, full polarization correlator providing at least 
      16,384 spectral channels per antenna pair (baseline). The correlator 
      provides full polarization capability for four polarization pairs of 
      input signals of up to 2 GHz bandwidth each.}

\item{A new real-time control system for the array and new monitor and control 
      hardware for the electronics system.}

\item{New high-level software that will provide ease of use of the EVLA to 
      its users.}

\end{itemize}

A simplified block diagram for the EVLA, showing the relationship between the 
major hardware and software sub-systems designed to meet these technical 
requirements, is shown in Figure~\ref{fig:blockdiagram}. Details of the 
subsystems are provided in the sections below, but at this point we note that 
six top-level constraints influenced the design of the EVLA sub-systems:

\begin{itemize}

\item{New systems must have a compatibility mode which allows continued 
      observations with them during the transition period when the old VLA 
      systems are still in use. This constraint particularly impacted the 
      monitor and control software which had to be compatible with the 
      original Modcomp computer-based system of the VLA and the new digital 
      data transmission system which had to provide a means of re-generating 
      the four, narrow band, IF signals required by the VLA correlator.}

\item{New systems must be designed to work with those parts of the old system 
      which were to be re-used. The dominant impact here was on the feed and 
      receiver subsystems where the decision to re-use the VLA antennas and 
      their optics forced the new feeds to be designed for the existing 
      subreflector illumination angle of $18^\circ$ (full angle).}

\item{The old analog IF transmission system was to be replaced with a digital 
      IF transmission system in order to provide improved bandpass stability 
      as well as an increased bandwidth. The bandwidth required to provide 
      digital transmission of 16 GHz of bandwidth per antenna required the 
      replacement of the original VLA waveguide system with fiber optics.}

\item{A design to minimize the impact of radio frequency interference (RFI) 
      was particularly important for two reasons. The decision to digitize 
      the IF signals at the antenna placed a large amount of high-power, 
      high-speed digital circuitry very close to the high sensitivity 
      receivers. Consequently, the feeds and receivers were designed with 
      shielding against RFI coming from within the receiver cabin, and the 
      electronics modules of the monitor and control, IF, LO and digital 
      transmission systems used components designed to minimize RFI. These 
      components were located inside custom-built, RFI-tight enclosures. 
      Additionally, the requirement to have continuous frequency coverage 
      in the receivers  from 1 to 50 GHz meant that the signals from a number 
      of strong sources of external RFI, such as radars and communication 
      satellites, are present in some of the receiver bands requiring a design 
      to minimize the impact of external RFI. Measures taken here included 
      high headroom in the receiver chains, up to 8 bits of resolution in the 
      IF digitizers, the provision of high spectral dynamic range in the
      correlator, and the option to include narrowband stop filters in the RF.}

\item{The software must all be loosely coupled, allowing access to all the 
      separate subsystems without interdependencies between them, so that 
      the entire system can continue to function although a particular 
      subsystem may not be functioning.}

\item{Information flow in the software system must occur with as little 
      human intervention as possible, to facilitate ease of use for users, 
      to minimize the manpower required to operate the system, and to 
      minimize errors in data transcription by people.}

\end{itemize}

\section{Major Subsystems}
\label{sec:subsystems}

\subsection{Front End Systems}

One of the major goals of the EVLA project is to provide continuous frequency 
coverage between 1 and 50 GHz from the secondary focus with an improved 
gain-system temperature quotient (G/T) and up to 8 GHz of instantaneous 
bandwidth in each of two, orthogonal, circularly-polarized channels. A number 
of novel wide-bandwidth technologies had to be employed to achieve these 
requirements.

The 1-50 GHz coverage is an increase of almost a factor of five over the 
coverage provided by the VLA and is broken up into the eight receiver bands 
shown in Table 1. Also listed are the goals for system temperature, antenna 
aperture efficiency, and effective system temperature. The system temperatures 
are based on an assumption of good weather and are mid-band estimates. Values 
at the band edges are typically degraded by up to a factor of two. 

\subsubsection{Feed Horns}

A new feed cone at the secondary focus of the telescope was designed to 
accommodate the eight feed horns and receivers. All the feeds are located 
near the vertex of the primary reflector on a circle of radius 97.54 cm 
centered on the reflector axis. An offset Cassegrain geometry is used so that 
the secondary focus lies on this circle. Changes in observing band are made 
by rotating the subreflector to redirect the reflected radiation to the 
desired feed horn. 

The octave bandwidth feeds for L, S and C-bands use compact profile corrugated 
horns [3]-[7]. In the compact horn design, the transition from the throat to 
the aperture uses a cosine taper. As a result, the horn is shorter in length 
by 30\% when compared to a linear taper horn, and the aperture of the horn is 
about 25\% smaller. A G/T analysis at 3 GHz determined that the optimum 
illumination taper at the edge of the subreflector for the EVLA antenna would 
be -17 dB. The limitation of space around the feed circle precluded the use of 
feeds large enough to accomplish this at the lower frequency bands. As a 
compromise, a taper of -13 dB was selected for S and C-bands resulting in 
smaller feeds with G/T reduced by only 10\%. For L-band, a taper closer to 
-10 dB was selected because the feed had to be about 20\% smaller in size to 
fit in its slot on the feed circle without requiring structural modifications 
to the antenna backup structure. This structural limitation is the reason for 
the lower aperture efficiency of the L-band receiving system (Table 1). The 
original VLA L-band feed is a hybrid, lens-corrected design, and has a higher 
G/T than the new design over its narrower 1.3-1.8 GHz frequency range at 
elevation angles above $60^\circ$. However, the new design has superior 
performance across its 1-2 GHz bandwidth below $60^\circ$ elevation due to a 
reduced forward spillover that gives less pickup from the ground. For the 
higher frequency bands (X, Ku, K, Ka and Q-bands), the feed designs are based 
on corrugated horns with a linear taper [8].

In order to obtain a good impedance match over these broad frequency ranges, 
especially the octave bandwidths needed at L, S and C-bands, ring-loaded 
corrugations are used in the mode converter section of each horn [9]. The mode 
converter provides for a good match between the corrugated and uncorrugated 
waveguides at the input of the receiver. The diameter at the last corrugation 
is small enough to prevent the excitation of the unwanted EH12 mode. The HE11 
mode in the corrugated section is converted to the TE11 mode in the circular 
waveguide. The mode converter, depending on the frequency band, has six to 
eight ring-loaded corrugations. 

All of the EVLA feed horns perform satisfactorily. Cross-polarized sidelobes 
are found to be below -25 dB. The measured return loss on the higher frequency 
horns is better than -25 dB. The lower frequency, compact horns exhibit a 
return loss better than -25 dB over most of their frequency range, dropping 
to about -18 dB at the low end of each band.

\subsubsection{Receivers}

\begin{figure}
\begin{center}
\epsfig{file=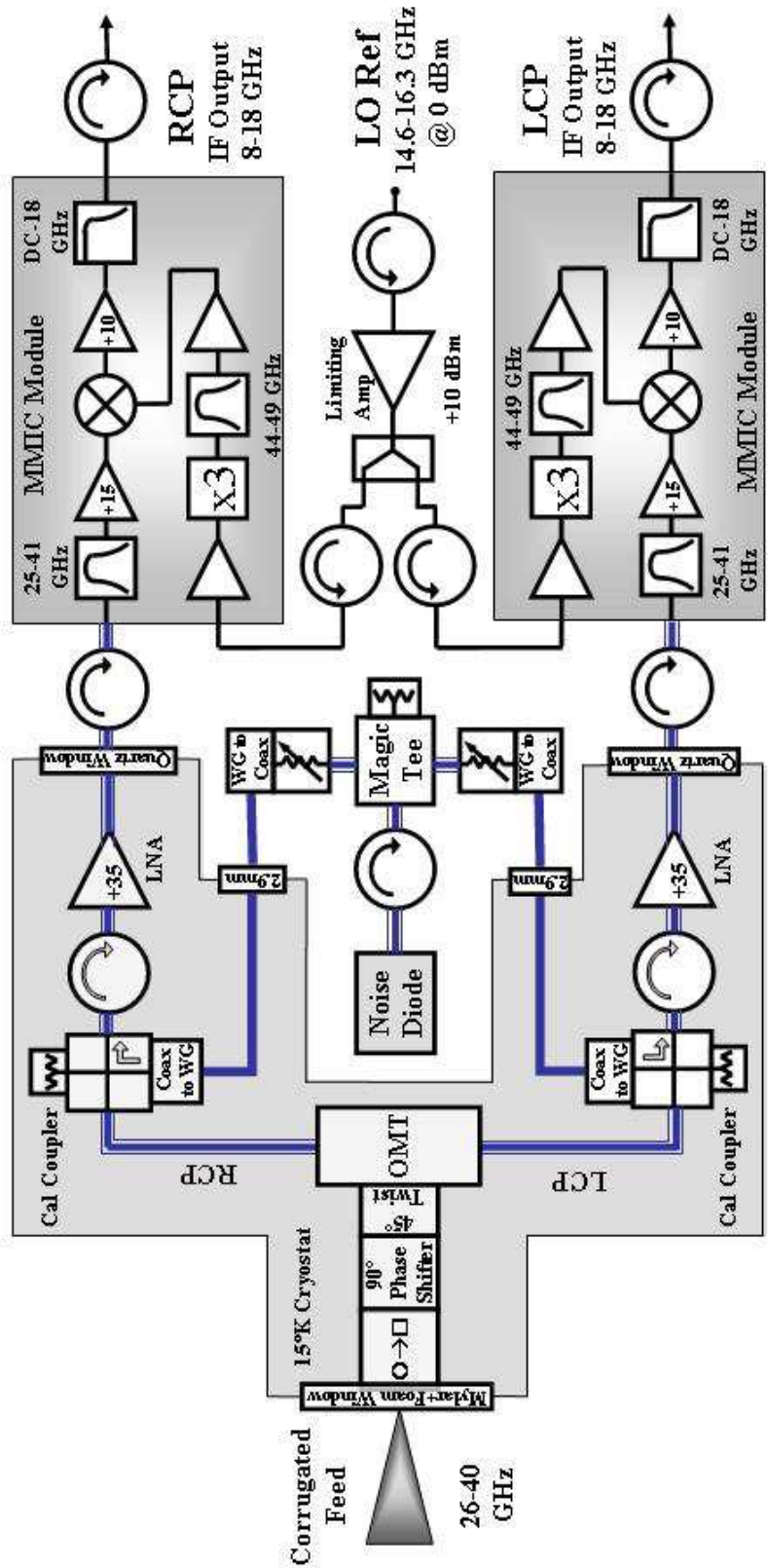, height=9in, angle=0}
\end{center}
\caption{Block diagram of the EVLA Ka-band receiver}
\label{fig:frontend}
\end{figure}

The EVLA project requires 240 new or upgraded receivers in total (8 receiver 
bands for 28 antennas and 2 spares). Existing VLA K and Q-band systems were 
upgraded with the latest low-noise cryogenic amplifiers and a new, broadband, 
RF/IF downconverter scheme to provide bandwidths up to 8 GHz. The new L, C, 
X and Ku-band receivers replace older VLA systems which were both narrowband 
and less sensitive. The S and Ka-band receivers are brand new systems. The 
existing 74 MHz (\lq\lq 4-band") and 327 MHz (P-band) receivers at the 
prime focal point of the antennas remain in place and are not modified as part 
of the project. A block diagram of the EVLA Ka-band receiver, which is similar 
in overall design to the devices used at the other EVLA receiver bands, is 
shown in Figure~\ref{fig:frontend}.

\subsubsubsection{Circular Polarizers}

All of the receivers accept dual, circularly polarized, signals. The L, S 
and C-band systems use quad-ridged orthomode transducers (OMTs) to separate 
the orthogonal linearly polarized components of the signal [10]. The design 
of an OMT with an octave bandwidth, a return loss better than -15 dB and 
which contains no trapped-mode bandpass features was a challenging development 
effort. This work was carried out at the National Radio Astronomy Observatory 
(NRAO). The mechanical design of the OMT also had to be one that was low cost 
and which could be easily fabricated and assembled in production quantities 
with minimal tuning adjustments. A further challenge was that the OMT needed 
to be small enough to be cryogenically cooled to reduce the noise contribution 
from inevitable ohmic losses. The OMT designs were extensively analyzed and 
optimized using theoretical models and finite-element electromagnetic 
simulation tools. Extensive parametric analyses were carried out to determine 
the most critical dimensions and to set fabrication tolerances.

The quad-ridge OMT must match the impedance of the circular waveguide presented 
by the antenna feed to the impedance of two, 50-ohm, coaxial transmission 
lines over an octave bandwidth. A square OMT structure was chosen instead of 
a circular structure because the single-mode bandwidth is wider and it is 
easier to manufacture. The high impedance, square waveguide port of the OMT 
tapers to a quadruple-ridge waveguide with an impedance near 50-ohms. A 
shorting structure in the quad-ridge waveguide behind the coaxial probes 
provides dominant-mode impedance matching while allowing in-band, higher-order 
modes to propagate through to the back of the device and into an absorber. 
This prevents trapped-mode resonances from occurring which would have otherwise 
resulted in sharp suck-outs in the frequency response, a common problem in 
quad-ridge OMTs.

Each quad-ridge OMT consists of two pairs of machined fins and identical cast 
or machined shells for the four corner pieces. The design of the OMT provides 
for repeatable fin spacings and allows each device to be easily tuned simply 
by adjusting the length of the coaxial probes. The L-band OMT is 69.62 cm long 
(corresponding to $2.32\lambda$ at the lowest frequency of 1 GHz) while the 
length of the C-band OMT is 21.40 cm ($2.85\lambda$ at 4 GHz). All the OMT 
designs have been found to perform well and meet or exceed specification. The 
S-band OMT, for example, has a measured return loss of less than -21 dB, an 
isolation better than -40 dB, and a warm insertion loss of -0.2 dB.

The right and left-circularly polarized components of the received radiation 
are separated in a combination of the quad-ridge OMT with a coaxial $90^\circ$ 
quadrature hybrid coupler. The signal paths between the OMT and the hybrid 
coupler are phase-matched to within $\pm 1^\circ$ in the center of the band. 
The amplitude balance between the two signal paths is better than 0.2 dB. The 
octave quadrature coupler is a commercially-available device that functions 
reliably at cryogenic temperatures. 

For the Ku, K and Ka-bands, the circular polarizers are pure waveguide devices. 
They use a corrugated, waveguide phase-shifter [11] followed by a 
twofold-symmetric, orthomode junction that was introduced by B{\rm\char'34}ifot 
[12]-[14]. The phase-shifter provides the $90^\circ$ phase-shift required to 
convert circular into linear polarization and typically has less than a 
$\pm 5^\circ$ phase error across the band. The symmetric OMT can be easily 
manufactured as a split-block with conventional machining techniques. The 
combined insertion loss of this style of polarizer is less than -0.3 dB across 
its operating bandwidth. The return loss is typically better than -20 dB and 
the isolation exceeds -40 dB. For the Q-band receiver with its relatively narrow 
bandwidth ratio (25\%), the circular polarizations are extracted with a 
commercially-available, sloping septum polarizer [15]. 

The development of an 8-12 GHz OMT for the X-band receiver is still in 
progress. Due to cryostat size and compressor capacity constraints, the ideal 
solution would be a new OMT that could be inserted directly into the existing 
VLA 8.0-8.8 GHz cryostat with minimal modifications. A planar OMT [16]-[18] 
may be the only design that will allow this option to be viable. This type of 
design replaces the coaxial probes with a four probe microstrip circuit that 
requires two $180^\circ$ hybrid couplers to combine the signals from opposing 
probes, as well as a $90^\circ$ hybrid to generate the left and right circular 
polarizations from the two orthogonal linear polarizations. While the benefits 
of the small size and weight of a planar OMT approach are obvious, it is yet 
unclear how its inherent resistive losses will affect its noise performance 
compared to a more traditional waveguide polarizer. Two types of planar OMT 
designs will be investigated - one using gold thin-film microstip circuits while 
the second uses high temperature superconductors. An all-waveguide OMT design 
is also being developed in the event that the noise contribution in the signal 
path prior to the low-noise amplifiers is excessive. This device is based on an 
ultra-thin turnstile design [19]. Its lateral dimensions would require a new 
cryostat to be designed to accommodate it.

\subsubsubsection{Low Noise Amplifiers}

There are nine different types of low-noise amplifiers (LNAs) used on the 
EVLA, all of which were designed and built by the NRAO's Central Development 
Laboratory [20]-[22]. The primary active elements in the amplifiers are indium 
phosphide, heterostructure, field effect transistors (InP HFETs) manufactured 
by TRW (now Northrop-Grumman Space Technology) under the NASA-led Cryogenic 
HEMT Optimization Program managed by the Jet Propulsion Laboratory, and 
obtained by the NRAO under special agreement. All the HFET devices used in the 
input stage of the EVLA amplifiers are from the so-called \lq\lq Cryo-3" wafer 
which exhibits record performance when compared to all other manufacturing runs 
of similar devices. In the 4-100 GHz range, the cooling of an InP HFET typically 
reduces the internal amplifier noise to within 4 to 6 times the quantum limit 
[23].  

The development of coolable LNAs requires detailed knowledge of both the 
signal and noise models of HFETs at cryogenic temperatures. These models have 
been developed with sufficient accuracy to achieve designs with optimal noise 
bandwidth performance [24],[25]. The design of the latest generation of 
amplifiers used for the EVLA was largely built upon the success of the LNAs 
created by the NRAO for the Wilkinson Microwave Anisotropy Probe project. The 
use of \lq\lq chip and wire" technology allows for the best performing HFET devices 
to be chosen from the wafer to be employed as the input stage of the amplifier, 
thus ensuring a superior noise figure while avoiding a  repeatability problem 
that is often observed in devices at cryogenic temperatures from different 
parts of the wafer or from different wafers. Single-ended amplifier designs are 
used for most of the frequency bands, with the C, X and Ku-band LNAs having 
three gain stages while the K, Ka and Q-band designs utilize four. 

Most of the receivers have cryogenic isolators located on the input of each 
LNA to provide an adequate input impedance match. This is necessary in order 
to diminish standing waves in the front-end optics as well as to reduce 
cross-channel leakage in the circular polarizer. Since octave-wide cryogenic 
isolators are not commercially available at L or S-bands, NRAO developed 
several balanced amplifier designs with input return losses better than -15 
dB. For L-band, the 1-2 GHz LNAs are built as \lq\lq gain blocks" with 18 dB 
of gain. Two cascaded gain blocks are used per channel (for a total of four 
amplifiers in each cryostat). The first, \lq\lq low-noise", balanced amplifier 
uses InP HFET transistors to achieve a noise temperature of about 4K. The 
second, \lq\lq high-power", gain block uses commercial, pseudo-morphic HFETs 
to achieve a 1 dB compression point in excess of $+13$ dBm, but with a higher 
noise temperature of about 20K. This configuration provides the best compromise 
between low-noise and high dynamic range. It also ensures that the broadband 
cryogenic amplifiers are unlikely to saturate from strong RFI before the 
signal reaches the warm post-amplifiers outside the cryostat. A novel 2-4 GHz 
balanced amplifier that uses a total of four InP HFET devices was developed 
for S-band. Each of the quadrature signal branches contains two gain stages 
operating in series. They are amplitude and phase matched to provide the 
desired gain and high input and output return loss performance. 
	
Each EVLA receiver has a gain of about 35 dB inside the cryostat with an 
additional gain of 25 dB or more in the room temperature RF section outside. 
This portion of the signal chain is relatively straightforward in most of 
the EVLA receivers and consists of commercial isolators, filters, 
post-amplifiers, and mixers.

Each of the new EVLA receivers is contained in its own cryostat which is 
cooled by a CTI Model 350 refrigerator. Model 22 units are used for cooling 
the existing X and Q-band receivers, which have the smallest cryostats of 
the eight receiver bands. The entire signal path, including the polarizers 
and the low-noise amplifiers, is cooled to 15K, except at L and S-bands 
where the large polarizers are cooled only to 50K.
	    
\subsubsubsection{MMICs}

Both the Ka and Q-band receivers make use of monolithic microwave integrated 
circuits (MMIC) in their room-temperature signal chain. The Department of 
Electrical Engineering at the California Institute of Technology was contracted 
to design a 40-50 GHz filtered post-amp for Q-band and a 26-40 GHz 
downconverter for Ka-band [26], [27]. These modules used custom 
metamorphic-HEMT low-noise amplifiers, which are InP pHEMT devices on a GaAs 
substrate, fabricated by Raytheon. The broadband downconverter for the Ka-band 
receiver also has a number of commercial MMIC amplifier chips, as well as a 
custom double balanced mixer and a LO tripler fabricated by United Monolithic 
Semiconductor using their GaAs Schottky diode technology. 

\subsubsubsection{Receiver Headroom}

In order to avoid the adverse effects of non-linearities which may arise in 
any of the active components (such as amplifiers, mixers or solid state 
attenuators) in the signal path when driven into compression, the EVLA 
front-ends and downstream IF circuitry were designed to ensure that the 
standard operating point is well below the saturation level of any device. 
This is a delicate balance since the signal must also be well above the 
noise floor of the various amplifiers in the RF/IF chain to ensure the 
overall system temperature is not degraded. The headroom specification for 
the EVLA is to have all active components operating at least 20 dB below their 
1\% compression points (which corresponds to being about 32 dB below the 1 dB 
compression point) when the antenna is looking at cold sky. The impact of 
insufficient headroom will only become a major limitation in the presence of 
strong RFI, where unwanted harmonics and intermodulation products generated in 
the amplifiers or mixers may cause spurious signals to arise. It is expected 
that the 20 dB headroom specification will provide adequate dynamic range so 
that the effects of both existing and future RFI will not cause saturation. 

\subsection{Intermediate Frequency and Local Oscillator Systems}

The LO and IF systems in the EVLA are highly flexible systems designed to 
support observations in two modes of operation.  During the early construction 
phase of the project, the EVLA is required to operate in a transition mode 
where the new electronics must operate in conjunction with unmodified VLA 
antennas and provide signals to the existing VLA correlator. Later in the 
project, as more VLA antennas are converted to the EVLA design and the WIDAR 
correlator is brought online, the systems will be able to easily switch between 
this transition mode and the final EVLA mode of operation. The transition mode 
will no longer be required once the original VLA correlator is decommissioned.

\subsubsection{Intermediate Frequency System}

\begin{figure}
\begin{center}
\epsfig{file=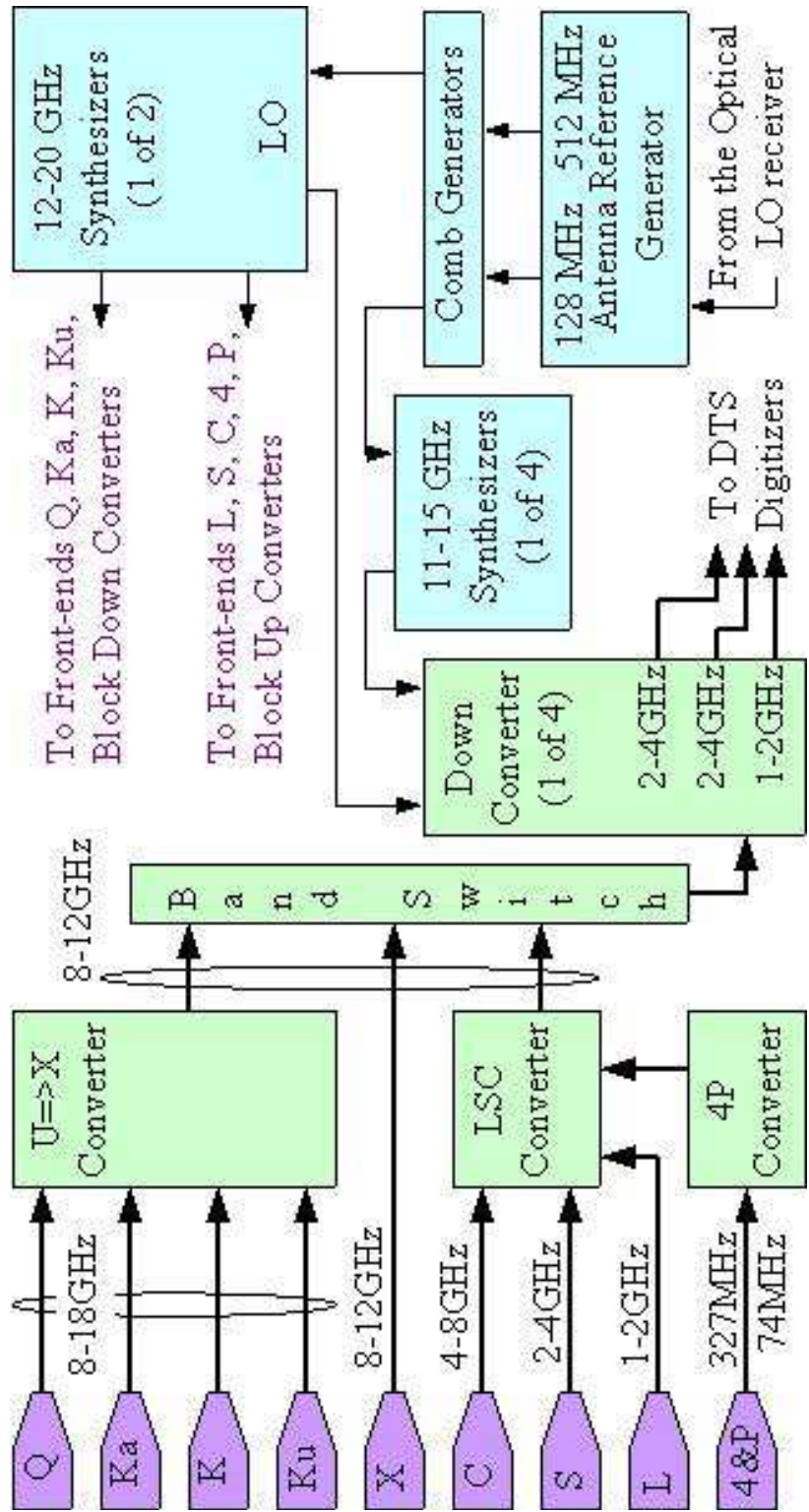, height=8in, angle=0}
\end{center}
\caption{EVLA antenna LO-IF system}
\label{fig:if}
\end{figure}

The EVLA IF system (Figure~\ref{fig:if}) consists of three primary signal 
paths: one for the higher frequency bands (Ku, K, Ka and Q), one for the lower 
frequency bands (4, P, L, S and C), and one for X-band.  All bands except X 
are converted to a common IF in the frequency range of 7.5-12.5 GHz.  The 
signals from the EVLA X-band receiver, which operates at 8 to 12 GHz, are 
routed directly into the main downconverters without frequency conversion. 

The RF signals from the high frequency receivers are converted to IF as 
follows. The first IF conversion for signals from the K, Ka and Q-band 
receivers occurs in the receivers themselves. At this stage of the IF, the 
frequency of the signals lies in the range of 8-18 GHz. IF conversion does 
not take place in the Ku-band receiver, which delivers its 12-18 GHz RF 
signal directly to the telescope's IF system. The signals from all four of 
the high frequency receivers are then routed to a broadband UX converter 
module. In this module, IF signals in the frequency range of 7.5-12.5 GHz 
are amplified and routed directly to the module's output. IF signals in the 
frequency range of 11.5-18 GHz are amplified and downconverted to a 7.5-12.5 
GHz IF using LO signals from the synthesizers discussed below.

Like the X and Ku-band receivers, no frequency conversion is performed in 
the 4, P, L, S, and C-band receivers. The 4 and P-band signals are combined 
and then routed to the 4P converter module where they are up-converted to 
the range of 1.0 to 1.4 GHz. These signals and those from the L, S and C-band 
receivers are routed to another conversion stage, in which the signals are 
up-converted to the 7.5 to 12.5 GHz IF frequency range, again using LO signals 
from the synthesizers described below.

Thus, signals from all receivers are converted to the 7.5-12.5 IF frequency 
range. In the final downconverter modules, the IF signal is bandpass filtered 
to 7.5-12.5 GHz and amplified. The amplitude of the IF signal is then leveled 
to a standard power by adjusting a 32-step attenuator. The leveled signal is 
then split into two paths.  Each of these paths is converted to a range of 
2048 to 4096 MHz using LO signals from a fine-tunable frequency synthesizer. 
These signals are routed into digitally controlled, 15 step, gain slope 
equalizers.  The equalizers are used to compensate for slopes in the signal 
bandpass due to variations in the receivers, cables and upstream electronics.  
Compensating for the slope is very important because of the limited dynamic 
range of the 3-bit digitizers (see below). The outputs of these equalizers 
are passed through a second 32 step attenuator to set the levels required by 
the digitizers. The signals are output from the module through very sharp 
cutoff, 2048-4096 MHz, anti-aliasing filters and connected to the 3-bit 
digitizers in the modules for the data transmission system (DTS) which 
operate on a 4096 MHz clock. Note that elsewhere in this document the IF 
bandwidth is referred to as a nominal 2 GHz band, although in fact it is 
2.048 GHz.

Additionally, one of the 2048 to 4096 MHz paths can be routed to an additional 
stage where it is converted by a 4096 MHz fixed LO signal to a range of 1024 
to 2048 MHz. This signal is leveled by a 32-bit step attenuator, then output 
from the module through a sharp cutoff, 1024-2048 MHz, anti-aliasing filter.
This is the input to the 8 bit digitizers in the DTS module, which operate 
on a 2048 MHz clock. Gain slope equalization is not performed on this signal 
due to the greater dynamic range of the 8-bit digitizer.   

\subsubsection{Local Oscillator System}

\begin{figure}
\begin{center}
\epsfig{file=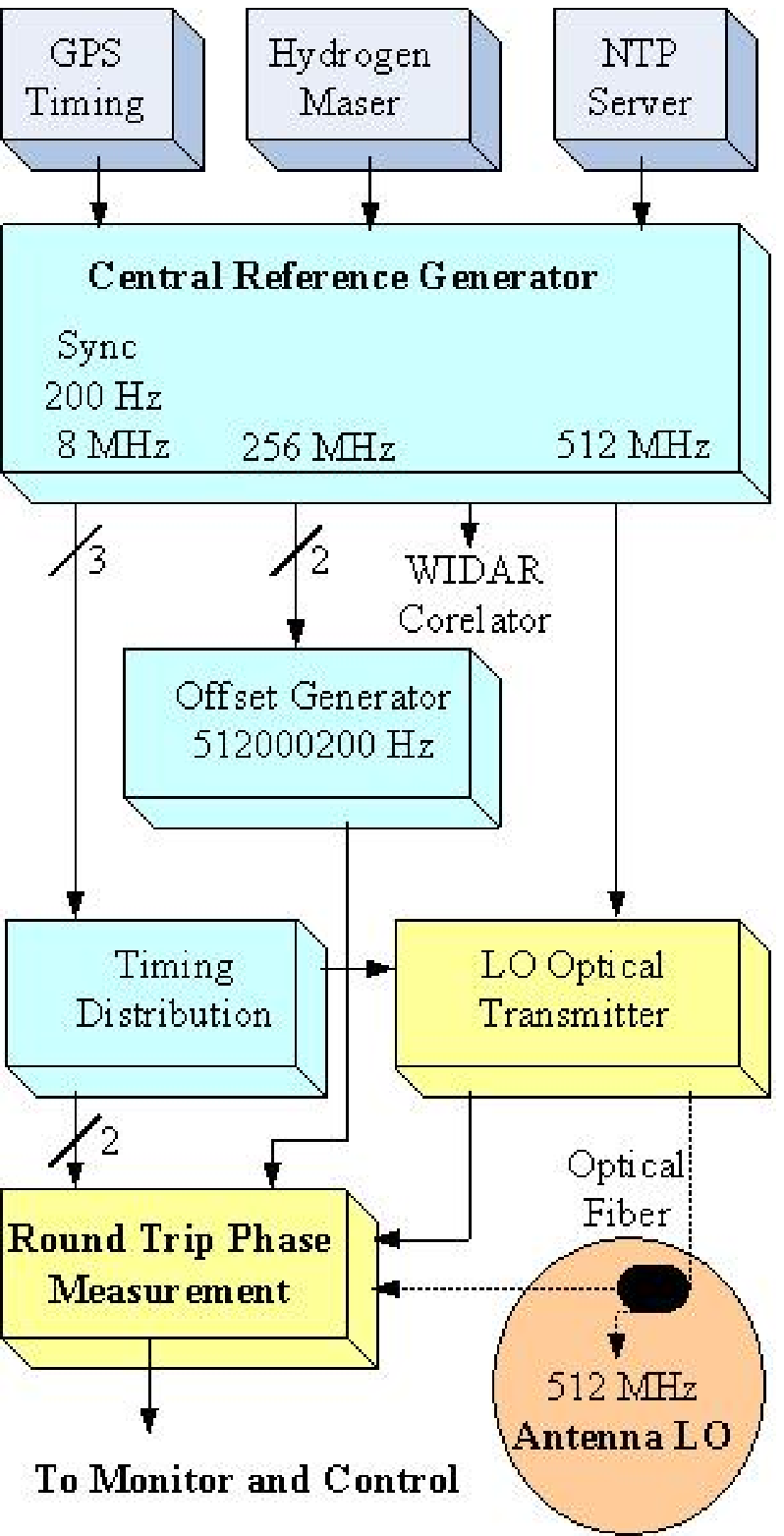, height=8in, angle=0}
\end{center}
\caption{EVLA LO distribution system}
\label{fig:lo}
\end{figure}

The local oscillator system (Figures~\ref{fig:if} and~\ref{fig:lo}) is based 
on 128 MHz and 512 MHz master reference signals, both from a crystal oscillator 
locked to a hydrogen maser and a 1 Hz, GPS-based, master timing signal. These 
signals are used to generate all other reference and timing information used 
by the EVLA system. With the exception of the maser, the master reference 
system is fully redundant to ensure high reliability and continued operations 
during maintenance.

Reference signals are distributed to an antenna over a single mode, fiber 
optic cable operating at the 1310 nm zero-dispersion wavelength. The reference 
signal consists of a 512 MHz sine wave with a phase inversion of 60 ns duration 
occurring once every 10 seconds as a timing reference. Absolute time is passed 
to the central and antenna electronics using the network time protocol (NTP) 
over the gigabit Ethernet employed by the EVLA's monitor and control system. 
Precise time is obtained in the antenna by synchronizing the NTP time with the 
timing signals in the local electronics. 

The phase of the reference signal to each antenna can change due to temperature 
and mechanical effects in the fiber. The change is measured by a round trip 
phase (RTP) system (Figure~\ref{fig:lo}). The measurement is made by splitting 
the optical reference signal at the antenna and sending part of it back to the 
central LO system on a second single mode fiber in the same cable. This signal 
is then compared to a master reference signal to get the total phase change in 
the fiber. The result is divided by two to get the phase change at the antenna.
Although there is no guarantee that the two fibers will behave exactly the 
same, experience indicates that they behave sufficiently similarly that the 
factor of two mentioned above is adequate for the needed accuracy. The accuracy 
of the phase measuring equipment is of order 100 fs, which is smaller than the 
relative delays introduced into the astronomical signals by the varying 
atmosphere above the antennas for most VLA baselines.

At the antenna, the optical signal is converted back to an electrical signal 
and routed to the antenna reference generator module (Figure~\ref{fig:if}). 
In this module, a crystal oscillator is phase locked to the incoming reference 
signal. The 128 MHz and 512 MHz signals from the oscillator are fed to two, 
step recovery diode-based, comb generators used to derive all of the higher 
frequency reference and LO signals in the antenna. In addition, a 60 ns timing 
pulse is extracted from the reference signal by a high speed flip-flop and 
fed into a field programmable gate array (FPGA). This FPGA develops all of 
the various timing signals used in the antenna.    

There are two types of synthesizers in the system (Figure~\ref{fig:if}). The 
LO signals to the receivers and/or converters for all bands except X-band are 
produced by a conventional Yttrium Iron Garnet (YIG) oscillator that is phase 
locked to a harmonic of the 512 MHz reference, offset by the 128 MHz reference, 
which produces tones in the frequency range of 11.776 GHz to 19.968 GHz in 
256 MHz steps.  Since the outputs of these synthesizers are used as the first 
(and in some cases second) LO, the synthesizers are designed to produce 
exceptionally clean, low phase noise CW signals with minimal harmonics.
This is especially critical for the K, Ka and Q band receivers where the LO 
is multiplied by an additional factor of two or three inside the receiver.     
 
The LO signals to the main downconverters are generated by the second type 
of synthesizers (Figure~\ref{fig:if}). These units are unique microwave synthesizers 
based on two YIG oscillators and two direct digital synthesizer (DDS) chips.
The synthesizers produce tones in the frequency range of 10.8 GHz to 14.8 GHz 
in microhertz steps. The design of this synthesizer was very challenging 
because the phase of its output signal was required to be controllable as 
the frequency is changed. The fine tuning requirement for the synthesizer 
was necessary for the transition mode operation of the EVLA system. In 
this mode, the very fine tuning capability of the synthesizer is used to 
produce a differential rate between antennas to allow the interferometer 
fringes to track the source. 

\subsection{Data Transmission System}
\label{sub:dts}

\begin{figure}
\begin{center}
\epsfig{file=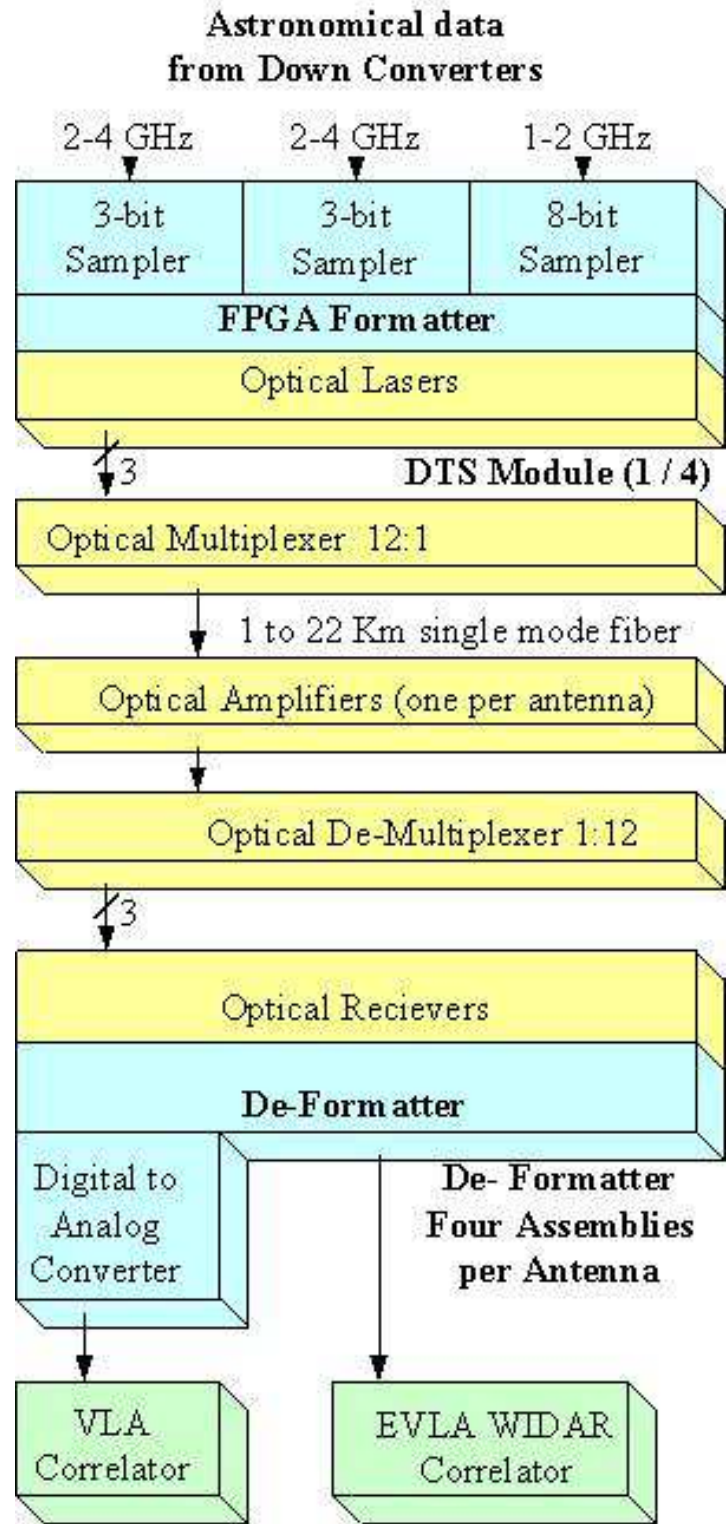, height=8in, angle=0}
\end{center}
\caption{EVLA data transmission system.}
\label{fig:dts}
\end{figure}

The EVLA data transmission system (DTS) digitizes the IF signals at the 
antennas and transmits them to the correlator located in the central control 
building. The DTS consists of a formatter module in the antenna, optical 
fibers between the antenna and the control building, and deformatters in the 
control building (Figure~\ref{fig:dts}). The deformatters also include signal 
processing circuitry and digitial-to-analog converters to allow the digital 
data produced by the EVLA antennas to be processed by the analog inputs to 
the existing VLA correlator during the transition period.

The data path from the antenna to the control building can be summarized as 
follows. An EVLA receiver can provide an instantaneous bandwidth of up to 8 
GHz in each of two polarizations. The receiver output is partitioned into 
eight, 2 GHz wide, IF bands by the LO-IF system (Figure~\ref{fig:dts} shows 
two of the eight bands). Each IF band can be sampled with 4 GHz samplers at 
3 bit resolution. Alternatively, the IF bands can be limited to 1 GHz bandwidth 
and sampled with a single, 2 GHz, 8 bit device for observations with a high 
dynamic range in the input signals. This latter configuration is normally 
used for observations at 4/P, L, S, and C-bands. The digitizers incorporate 
a de-multiplexer that reduces the data rate to 512 Mbps on a parallel data 
path. A formatter receives the data from the digitizers and prepares them for 
shared transmission over three OC-192 fiber optic links at 10 Gbps each. The 
deformatter in the control building receives the three fiber optic signals 
and reorganizes the data into a format that is recognized by the new EVLA 
correlator.

The two 3-bit digitizers, the 8-bit digitizer, and the digital formatter are 
contained in a formatter module, and each antenna has a total of four formatter 
modules. The 8-bit digitizer is a 10 bit, analog-to-digital converter with a 
de-multiplexer chip set that is commercially-available from e2v. The 3-bit 
digitizer uses a 6-bit, 6 Gsps chip from Teledyne Scientific. The high speed 
de-multiplexing function is performed by serial-to-parallel converters from 
On Semiconductor. The formatter assembly consists of three FPGAs and three 
custom transponders. The FPGA receives data from the digitizers and splits 
the data stream into 128-bit payload frames. The FPGA also adds framing, 
timing, and error detection information. Since the EVLA DTS is unidirectional, 
it uses the specially constructed transponders to transmit the data on to 
three high-speed optical fibers. The data are transmitted on a fiber at 10 
Gbps using 1550 nm lasers. The three data channels are completely independent 
apart from synchronization, monitor, and control signals, but all three are 
required to reproduce the sampled signal. The output data from the four 
formatter modules (i.e. 12 signals at 10 Gbps each) are combined onto a single 
fiber using dense wavelength division multiplexing (DWDM). This arrangement 
requires a different wavelength laser in each transponder.

A set of fibers runs to each of the 72 antenna pads throughout the EVLA. The 
cables are set in a star configuration with all fibers originating at the 
control building and ending at each antenna pad. The fibers for each antenna 
are grouped together in trunks and are terminated at each pad. 

At the control building, the DWDM signal from each antenna is optically 
amplified and de-multiplexed onto 12 separate fibers. Each deformatter receives 
a set of three fiber signals from one of four formatter modules in an antenna 
and reorganizes the data into the format required by the correlator. The 
deformatter is physically installed in the correlator as a daughter card on 
the correlator's station board. (Architecturally, it is considered part of the 
antenna's electronics, instead of a component of the correlator). Each of 
these three signals is processed by an optical receiver and an FPGA. The 
optical receiver converts the optical input to a 10 Gbps electrical signal, 
which is de-multiplexed to a 640 Mbps data rate that can be accommodated by 
the FPGA. Logic in the FPGA identifies frame boundaries in the input data 
stream and performs error detection and reporting with a simple parity check 
scheme. A 5-bit frame counter in the input frame format is used to synchronize 
the three channels. The data are then organized into the format required by the 
correlator.

A major requirement of the project is to continue scientific observations with 
the VLA as the EVLA electronics systems are installed. To accommodate this 
transition requirement, a digital filter was implemented in the deformatter 
FPGAs that selects a 64 MHz sub-band from the 1 GHz bandwidth available. The 
sub-band is converted to analog and passed on to the VLA baseband system (see 
Figure~\ref{fig:dts}). 

Digitization at the antenna avoids the distortions contributed by an analog 
transmission system, but raises concerns regarding self-generated RFI. 
Therefore, several features were implemented in the formatter module to 
suppress RFI. Packaging of the digitizers and associated electronics into a 
single formatter module avoids the routing of high-speed digital signals 
between modules and shortens signal runs. Clock signals and RF digitizer inputs 
enter the module via coaxial cables bonded at the module wall. Power to the 
module is provided by a single, heavily filtered, 48 volt supply. All operating 
voltages for electronics inside the module are obtained from regulators 
internal to the module. The high speed digital output data exit the module via 
three single-mode optical fibers. All fiber-optic signals penetrate the module 
wall through connectors chosen for their RF attenuation. The penetrations act 
as waveguides beyond cutoff and provide good shielding characteristics. The 
high-speed electronics in the module are cooled by air flowing through 
honeycomb RFI filters on the top and bottom of the module enclosure. The 
formatter module is heavily shielded. The main module enclosure consists of 
a welded aluminum box with one open end and the honeycomb filters attached on 
its top and bottom by screws with absorbing RF gaskets. The major electronics 
assemblies are attached to a single aluminum bulkhead that runs down the 
center of the module. Final assembly of the module consists of sliding the 
bulkhead into the box and securing its front cover with an absorbing RF gasket 
and a large number of screws. The shielding effectiveness of the entire 
assembly has been measured at 85 dB.

\subsection{Correlator}

The EVLA WIDAR (Wideband Interferometer Digital Architecture) correlator is 
the final destination for all of the real-time wideband signals carefully 
collected, down-converted, and sampled by all of the antennas. It calculates 
the cross-correlation function for every pair of antennas (baseline) in the 
array. This is no small feat considering the number of antennas in the EVLA 
(27), the bandwidths observed (16 GHz per antenna), and the $10^3$ to $10^6$ 
spectral channels needed per baseline.  

The correlator is fundamentally an XF-type correlator, which has its roots 
in, and shares signal-processing elements with, a correlator [28] developed 
for the VSOP space radio telescope.  An XF correlator cross-multiplies data 
from different antennas prior to the Fourier transformation to the frequency 
domain, as opposed to an FX correlator where the Fourier transformation 
precedes the cross-multiplication. More details on the fundamental signal 
processing for WIDAR are described in [29]. The correlator is sometimes 
referred to as an FXF style correlator, wherein the wideband signal is divided 
into smaller sub-bands with digital filters. Each sub-band is subsequently 
correlated in time and Fourier-transformed to the frequency domain. The sub-bands 
can be \lq\lq stitched" with others to yield the wide-band cross-power result. 
Aliasing at the sub-band edges is greatly attenuated using offset LOs [29] in 
the antennas. The LO offsets also perform the equivalent of the Walsh function 
phase switching that is currently used at the VLA. Frequency offsets are 
typically 1 kHz, with a minimum of 100 Hz.

The primary feature of the EVLA WIDAR correlator is the large number of 
independently tunable sub-bands that are produced by digital filters implemented 
in FPGAs. The correlator is also robust to RFI given its large number of bits per 
sample and its filter reject-band attenuation. Each sub-band is tunable in location 
and bandwidth within the 2 GHz-wide basebands of the EVLA and can be assigned a 
flexible number of spectral channels. Tradeoffs can be made for bandwidth, number 
of spectral channels per sub-band, and field-of-view on the sky. The sub-band 
reject-band attenuation is better than 60 dB. 16K to 4M spectral channels can 
be produced per baseline, and the expandable 32-antenna correlator can process 
496 baselines, each with 16 GHz total bandwidth. The correlator also contains 
high-performance pulsar \lq\lq phase-binning" for \lq\lq stroboscopic"
imaging of pulsars. Additionally, it can produce the phased-array sum on 
the entire 16 GHz bandwidth of the array. The phased array mode is used primarily 
for producing data for very long baseline interferometry (VLBI) and pulsar 
observations.

\begin{figure}
\begin{center}
\epsfig{file=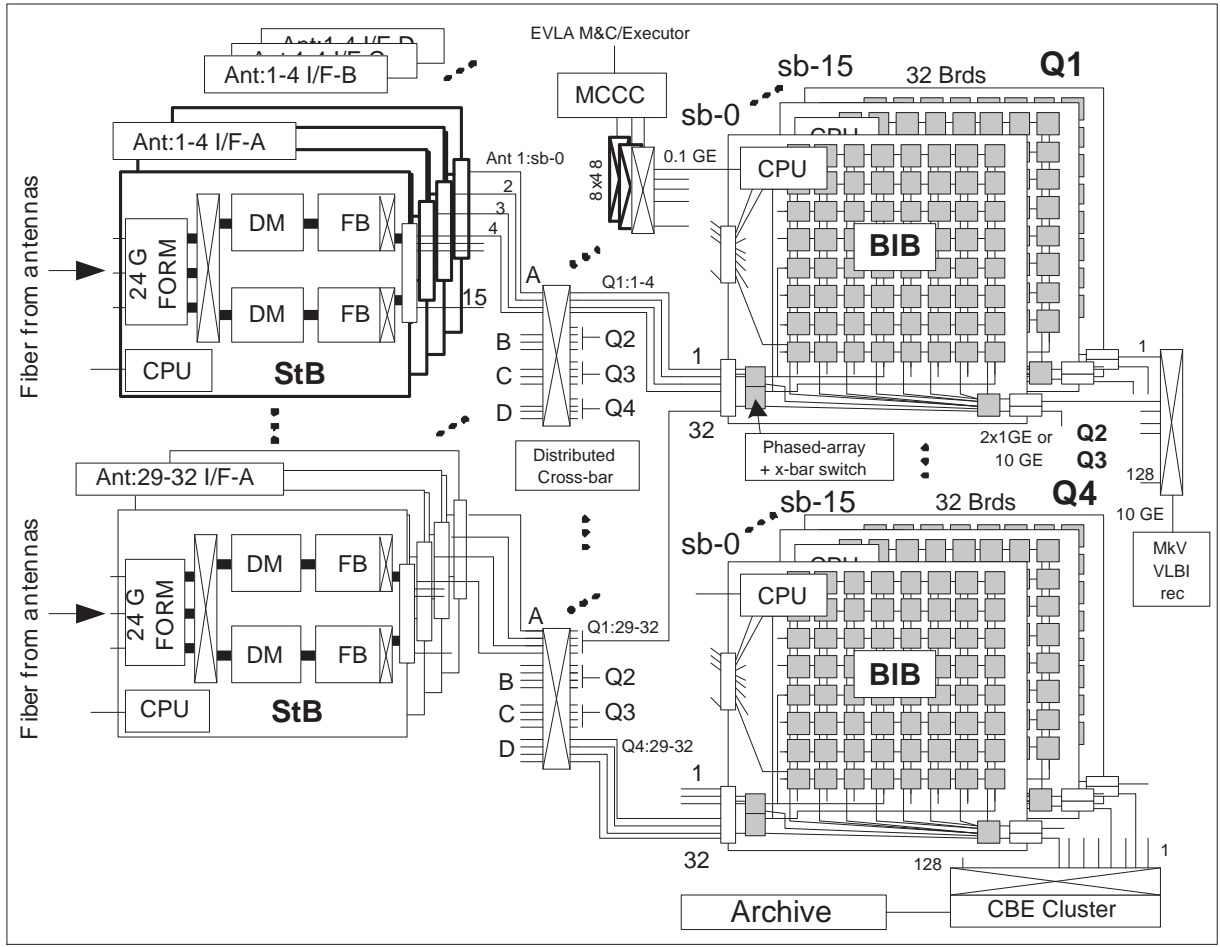, height=5in, angle=0}
\end{center}
\caption{Simplified block diagram of the WIDAR correlator.}
\label{fig:widar}
\end{figure}

The correlator consists of 16 standard 61-cm racks, each containing 16 large 
(38 cm x 48 cm) 28-layer, controlled-impedance, circuit boards of two types. A 
simplified diagram of the correlator, showing all key elements, is shown in 
Figure~\ref{fig:widar}. The 128 station boards (StB) and 128 baseline boards 
(BlB) are connected by a distributed cross-bar switch to provide the flexibility 
described above. The signals traveling between the boards are 1 Gbps LVDS/PCML. 
A total of 512 high-speed data cables connect the racks, with each cable carrying 
approximately 10 Gbps of data and control and timing information.  Buffers, 
embedded synchronization codes, and phase lock loops in FPGAs are used to 
eliminate the need for synchronization of clocks between boards or racks. Some 
minor restrictions apply compared to what a full cross-bar could accomplish; 
however, the cost and complexity of the distributed cross-bar are greatly 
reduced in comparison. 

The WIDAR signal processing path is as follows. 96 Gbps of sampled payload data 
arrive from each antenna via optical fiber. As described in subsection~\ref{sub:dts}, 
the data are sent to 4 StBs per antenna via the deformatter (the fiber optic 
receiver module (FORM) shown in Figure~\ref{fig:widar}). An FPGA cross-bar allows 
these signals to be routed to two 64-bit wide, 256 MHz data paths on the board. 
Coarse delay tracking to within $\pm$one-half a sample is implemented with an FPGA 
and DDR SDRAM on a delay module (DM) mezzanine card on the StB. Baseline delay at 
the level of $\pm$one-sixteenth a sub-sample is implemented by continuous, real-time 
tracking of the residual delay error at the center of each sub-band using the 
correlator chip's phase rotators [29]. The delay-corrected signals are then 
distributed to dual, 18-FPGA filter banks (FB). Each FB FPGA can further 
delay-correct the data to allow for a sub-band-specific beam delay-center offset 
on the sky. Each FPGA also contains four, 512-tap FIR stages with 16-bits carried 
between stages. One or all of the stages can be used to produce sub-bands from 128 
MHz wide to as narrow as 31.25 kHz. Data exit the StB into the distributed cross-bar 
and is then routed to the BlBs. High-density, hardmetric connectors and mating 
cabling allow for 1.024 Gbps per pair of BlBs, fitting the numerology of the 
correlator nicely. The pair of BlBs processes all 32 antennas (496 baselines) for 
a sub-band pair (2 x 128 MHz), with 64 Gbps of astronomical data and up to 32 Gbps 
of control and timing data into each BlB. Signals are synchronized using dynamic 
phase alignment, re-timed, and distributed by two cross-bar and phasing FPGAs to 
the 8x8 array of XF correlator chips. The F672 BGA XF correlator chip is a 4 Mgate 
standard-cell device fabricated in 130 nm CMOS. Each device contains 2048 
\lq\lq lags", each of which consists of a 4-bit multiplier, a 3-level phase rotator, 
and dual 22-bit accumulators. The phase rotators perform phase tracking (fringe 
rotation), sub-sample delay correction, and frequency shift removal. All signals 
for integration control originate in the StB, and integrations can be synchronized 
to system timing or any timing model, such as a pulsar ephemeris. Data from each 
correlator chip are integrated further in a low-cost LTA FPGA and DDR SDRAM. The 
final results are contained in UDP/IP packets, each containing 128 complex-lag 
accumulation results. The packets are sent to the correlator computing cluster on 
1 Gigabit Ethernet via a commercial switch. The cluster carries out the data 
normalization, Fourier transform, and interference excision. The resultant data 
are sent to an archive for final image processing.  

The crossbar FPGAs also perform the phasing of the antenna array. The phased 
signal is output to several destinations, primarily Ethernet packets to a VLBI 
data recorder.  

The correlator also provides four, wideband, auto-correlation products for every 
baseband pair. Each data product has 1024 spectral channels, but with a factor 
of four sensitivity loss due to hardware limitations that arise from acquiring the 
auto-correlations in 64-lag chunks every 10 ms. Auto-correlations are also 
acquired in each sub-band with the cross-correlator hardware, although only two 
products at a time per antenna may be acquired.

Each StB and BlB contains a PC/104+ embedded processor mezzanine card running 
real-time Linux. PC/104+ is an inexpensive and high performance technology for 
this purpose. The StB computing requirements are the most stringent because 
real-time delay tracking and phase model generation, integration control, and 
data acquisition must be performed. The BlB computing requirements are less 
stringent, requiring only monitor and control operations since the flow of all 
correlation coefficient data is CPU-independent. This arrangement effectively 
eliminates any bottlenecks to the data flow.  

The data rate produced by the correlator is governed by several factors. The 
correlator hardware is capable of producing data at an extremely rapid rate, so 
data rate limitations are set by a combination of factors, including configuration 
of the CBE computers, their performance and network topology, and not least 
scientific necessity. The standard correlator configuration is a 1 Gigabit 
Ethernet link from each of 128 BlBs for the visibility data, providing the 
capability for dumping all spectral channels every 11 ms, and a 1 Gigabit Ethernet 
link from each BlB for real-time phased data. Thus, the maximum data rate to the 
computing cluster is 128 Gbps. A BlB can be upgraded with 10 Gigabit Ethernet for 
a maximum data rate of 1.28 Tbps. With the planned number of CBE computers, all 
spectral channels can be dumped every 100 ms, for a data rate of 167 
mega-visibilities per second in a 32-station correlator. The actual dump times used 
will be determined by science objectives, array configuration, and data storage 
capabilities. On the timescale of early 2012, typical data rates produced by the 
correlator are expected to be at the more modest rate of about 20 MBps, with a 
maximum rate of 75 MBps.

\subsection{Software}

One of the major goals of the EVLA project was to replace the software of the VLA, 
much of which was written over 20 years ago and was becoming more and more 
difficult to maintain. The new hardware of the EVLA requires much more software 
to control it, and because the instrument is so much more capable than the VLA, 
the control and user software is necessarily more complex. While more complex, 
the intent of the new software in the end is to make the process of observing 
with the EVLA much more accessible to the astronomer who is not necessarily 
intimately familiar with the theory and techniques of radio interferometry.

The EVLA software is divided into two major areas: the control and monitoring of 
the hardware, or Monitor and Control (M\&C), and the software generally accessed 
by the user, or User Software. Figure~\ref{fig:blockdiagram} shows a very high 
level view of the EVLA software system, along with how it fits in with the hardware, 
and with the information flow highlighted. Two overriding principles guide the 
implementation of the software. First, the software must be loosely coupled, so 
that if a particular subsystem is not functioning, the rest can continue to run. 
Second, the software must make it easier to use and operate the EVLA, including 
lack of human intervention where possible.

\subsubsection{Monitor and Control Software}

The EVLA monitor and control software design concept is a highly distributed, 
loosely coupled system. Most hardware modules contain a fairly capable, single 
board computer system, the Monitor/Control Interface Board, or MIB (for the 
correlator, called CMIBs, for Correlator MIB). The MIBs receive commands via 
datagram (under the UDP protocol) from the central system, and send monitor data 
by datagram to a central location, which archives it in a general purpose database. 
The MIBs also support a telnet protocol, which provides a useful debugging 
interface.  Commands sent to devices/MIBs are labeled with a time at which they 
are to be executed. They are typically sent to the devices several seconds in 
advance, so the central entity itself does not, normally, have serious real-time 
deadlines to meet. Because self-generated RFI is a major problem for a sensitive, 
broad band radio receiving system, a central consideration for the MIB design was 
the minimization of generated RFI. For that reason, a system-on-a-chip was chosen, 
rather than the more common design with memory on an adjacent chip. MIBs also 
perform limit checking on the monitor points available to them. If a monitor point 
falls outside the range of permitted values, an alert is sent by datagram. These 
alerts are classified as to whether they are informational, require telescope 
operator action, or cause the data to be flagged as bad.

There are six central elements of the overall control system:

\begin{itemize}

\item{The software on the MIBs themselves, which makes the individual modules 
      within the system rather intelligent.}

\item{The observing general manager, called the Executor. Within the Executor is 
      a set of objects which are in one-to-one correspondence with the physical 
      antennas. These objects are the mechanism by which all commands are sent 
      to the MIBs in the antennas, and contain the code which converts a more 
      generic description of the desired observation setup into commands to the 
      MIBs associated with the antenna. The Executor operates on scripts, which 
      allows for full scripting control by the observation setup software or 
      astronomer.}

\item{The entity which receives broadcast monitor data from the MIBs and elsewhere. 
      It stores data in a general purpose database (currently Oracle). Web-based 
      retrieval tools are provided to extract and list or plot the data. There are 
      about 300 monitor points on each antenna, and about 6 million rows per day 
      are stored in the database.}

\item{The alert server receives the alerts generated by the MIBs and other parts 
      of the system, and classifies them as described above.}

\item{The entity which annotates data with the system state and setup is called 
      MCAF (Metadata Capture And Format). It gets data from the Executor (via 
      XML records broadcast by the Executor), from the alerts-server (for 
      information about whether data is valid for each antenna), and from other 
      parts of the system.}

\item{The entity which analyses output data in real-time is called Telcal. 
      Real-time analysis is required for correcting antenna pointing and for 
      solving for phases for use of the VLA as a VLBI element by coherently 
      summing the signals from the antennas. It is also very useful to have 
      it report system sensitivities during observations of calibrator sources 
      as a general, telescope operator-accessible indication of the system health.} 

\end{itemize}

In addition, the correlator (WIDAR) requires several major software elements:

\begin{itemize}

\item{CMIB software, which is similar to MIB software, allows even the lower 
      level hardware modules to be relatively intelligent.}

\item{The Correlator Back-End (CBE) software, which collects the data from the 
      station and baseline boards and combines them into the fundamental data 
      format for storage, the Binary Data Format (BDF). The CBE software does 
      the Fourier transform from lags to spectra, spectral windowing, data 
      blanking, sub-band stitching, and other tasks.}

\item{The software on the main correlator control computer, which allows for 
      communication with the rest of the EVLA software system. This communication 
      is all done via documents in the eXtensible Markup Language (XML) which 
      conform to a particular protocol (the Virtual Correlator Interface, or VCI).}

\item{The software controlling the power distribution to the various parts of the 
      correlator hardware. Although this software is relatively simple, it deserves 
      mention because power control within the system is so important.}

\end{itemize}
 
As mentioned, the system as a whole is loosely coupled. System components 
communicate via datagrams (sending either an internally developed command format 
or XML documents) or via the http-based REST protocol.  Any system component can 
be stopped and restarted with minimum impact on the rest of the system, and normal 
operation will return shortly after a system component is restarted, as soon as 
a set of commands to that component has been sent.

\subsubsection{User Software}

The EVLA user software encompasses all software that the user directly interacts 
with, as well as software that impacts the scientific output of the instrument. 
The user software is designed to be easy to use, with modern graphical user 
interfaces (GUIs) where needed. The user software, like the M\&C software, is 
designed to be modular, so that major subsystems do not depend on each other to 
run. Information is passed between subsystems either via storage to and retrieval 
from a database, or direct passage of XML documents. This makes it so that 
transcription of information by hand, either written or computer, is not necessary.

There are five main subsystems of the user software:

\begin{itemize}

\item{The entity for proposal creation, editing, submission, and handling, called 
      the Proposal Submission Tool (PST). This tool allows potential users to 
      propose for observing time on the EVLA, and for the evaluation of those 
      proposals by the observatory.}

\item{The entity for creation of observing scripts which give the details of 
      sources to be observed, the instrumental setups, and timing information, 
      called the Observing Preparation Tool (OPT). The output from this tool is 
      a Jython script to be input into the aforementioned Executor.}

\item{The entity for deciding which of all possible observing scripts available 
      for observing should actually be observed at a given point in time, called 
      the Observation Scheduling Tool (OST).  This tool takes into account the 
      scientific priority of the scripts, the required and current observing 
      conditions, and other constraints. This tool allows for observations to
      eventually be fully dynamically scheduled, i.e., no human intervention 
      involved.}

\item{The entity for access to data in the scientific archive, called the Archive 
      Access Tool (AAT).  This allows scientists to find their own proprietary data, 
      or other public data, given any of various search criteria. Once found, data 
      can be downloaded to the user computer, or sent to a data processing pipeline 
      for reduction, and the results from that downloaded.}

\item{The data post-processing software. This software element takes the fundamental 
      measured interferometric quantity, the visibilities, and processes them into 
      final images or spectra (or image cubes, the combination of the two). It 
      includes data editing, flagging, calibration (bandpass, flux density scale, 
      complex gain as a function of time, etc.), and imaging, including 
      self-calibration. Automatic data-processing pipelines are part of the more 
      general post-processing software, intended to be initially developed by 
      astronomers and then implemented within the package. All observations done in 
      standard modes will be pipeline processed and the results placed in the 
      archive along with the raw and calibrated visibility data.}

\end{itemize}

This user software must be capable of supporting both expert users and those that 
are new to using the EVLA, or any radio interferometer. As such, it often has two 
interface styles, expert and novice. In most cases, these interfaces are GUIs, 
either web-based, or stand-alone, using modern software tools (for instance, most 
of the web-based GUIs are written within the JavaServer Faces (JSF) framework).

\section{Summary} 
\label{sec:summary}

The EVLA is a major expansion to the highly flexible and productive VLA. The 
expansion includes new or upgraded receivers that enable continuous frequency 
coverage from 1 to 50 GHz, a new broadband LO/IF system, a new fiber optic-based 
data transmission system, a new correlator to process the wideband data, a new 
monitor and control system, and new software that provides telescope ease of use. 
The expansion provides order of magnitude, or greater, improvements over existing 
capabilities with the VLA. Observations with the VLA have been ongoing as the 
expansion has progressed. The project is scheduled for completion in 2012. The 
expansion will enable new investigations into celestial radio transients, the 
evolution of objects in the universe, and the structure and strength of celestial 
magnetic fields.

\acknowledgements

The National Radio Astronomy Observatory is a facility of the National Science Foundation 
operated under cooperative agreement by Associated Universities, Inc. The Dominion Radio 
Astrophysical Observatory is a National Facility operated by the National Research Council 
Canada.

\clearpage

\begin{center}
\bigskip
{\bf Glossary of Acronyms}
\end{center}

\begin{tabbing}
DDR SDRAM  \=  Double Data Rate Synchronous Dynamic Random Access Memory \kill
AAT  \>  Archive Access Tool \\
BDF  \>  Binary Data Format \\
BGA  \>  Ball Grid Array \\
BlB  \>  Baseline Board (in the WIDAR correlator) \\
CBE  \>  Correlator Back-End \\
CMIB \>  Correlator Module Interface Board \\
CMOS \>  Complementary Metal Oxide Semiconductor \\
CPU  \>  Central Processing Unit  \\
DDR SDRAM  \> Double Data Rate Synchronous Dynamic Random Access Memory \\
DM   \>  Delay Module (in the WIDAR correlator) \\
DTS  \>	 Data Transmission System \\
DWDM \>  Dense Wavelength Division Multiplexing \\
EVLA \>	 Expanded Very Large Array \\
FB   \>	 Filter Bank (in the WIDAR correlator) \\
FIR  \>  Finite Impulse Response \\
FORM \>  Fiber Optic Receiver Module \\
FPGA \>	 Field Programmable Gate Array \\
GaAs \>	 Gallium Arsenide \\
Gbps \>	 Gigabit per second \\
GHz  \>  GigaHertz \\
GPS  \>	 Global Positioning System \\
G/T  \>	 Gain-system Temperature quotient \\
GUI  \>	 Graphical User Interface \\
HEMT \>  High Electron Mobility Transistor \\
HFET \>  Heterostructure Field Effect Transistor \\
Hz   \>	 Hertz \\
InP  \>  Indium Phosphide \\
IF   \>  Intermediate Frequency \\
IP   \>  Internet Protocol \\
JSF  \>  JavaServer Faces \\
LNA  \>  Low Noise Amplifier \\
LO   \>  Local Oscillator \\
LTA  \>  Long Term Accumulator \\
LVDS \>  Low Voltage Differential Signaling \\
kHz  \>  kiloHertz \\
M\&C \>  Monitor and Control \\
MCAF \>	 Metadata Capture and Format \\
MBps \>  MegaByte per second \\
MCCC \>	 Main Correlator Control Computer \\
MHz  \>  MegaHertz \\
MIB  \>  Module Interface Board \\
MMIC \>  Monolithic Microwave Integrated Circuits \\
NASA \> National Aeronautics and Space Administration \\
NRAO \>	 National Radio Astronomy Observatory \\
NTP  \>  Network Time Protocol \\
OMT  \>  Orthomode Transducer \\
OPT  \>  Observation Preparation Tool \\
OST  \>  Observation Scheduling Tool \\
PCML \>  Program Call Markup Language \\
PST  \>  Proposal Submission Tool \\
REST \>  REpresentational State Transfer \\
RF   \>  Radio Frequency \\
RFI  \>  Radio Frequency Interference \\
RTP  \>  Round Trip Phase \\
StB  \>  Station Board (in the WIDAR correlator) \\
Tbps \>  Terabits per second \\
UDP  \>  User Datagram Protocol \\ 
VCI  \>  Virtual Correlator Interface \\
VLA  \>	 Very Large Array \\
VLBI \>  Very Long Baseline Interferometry \\
VSOP \>  VLBI Space Observatory Program \\
WIDAR \> Wideband Interferometer Digital ARchitecture \\
XF   \>  Cross multiplication/Fourier transformation \\
XML  \>  eXtensible Markup Language \\
YIG  \>  Yttrium Iron Garnet 
\end{tabbing}

\begin{deluxetable}{ccccccc}
\tablenum{1}
\tablewidth{490pt}
\tablecaption{Required Performance of EVLA Frequency Bands} 
\tablehead{
 \colhead{} & \colhead{Center} & \colhead{Frequency} & \colhead{System} & 
 \colhead{Aperture} & \colhead{Effective} & \colhead{Maximum IF} \\
\colhead{Band} & \colhead{Frequency} & \colhead{Range} & 
\colhead{Temperature} & \colhead{Efficiency} & \colhead{Temperature} &
\colhead{Bandwidth} \\
\colhead{} & \colhead{(GHz)} & \colhead{(GHz)} & \colhead{(K)} & \colhead{} &
\colhead{(K)} & \colhead{(GHz)} \\}
 
\startdata
 L & 1.5 &  1-2	& 26 & 0.45 & 58 & 2x1 \\
 S & 3.0 &  2-4 & 29 & 0.62 & 47 & 2x2 \\
 C & 6.0 &  4-8 & 31 & 0.60 & 52 & 2x4 \\
 X & 10	&  8-12	& 34 & 0.56 & 61 & 2x4 \\
 Ku & 15 & 12-18 & 39 &	0.54 & 72 & 2x6 \\
 K & 22 & 18-26.5 & 54 & 0.51 & 106 & 2x8 \\
 Ka & 33 & 26.5-40 & 45 & 0.39 & 115 & 2x8 \\
 Q & 45	& 40-50 & 66 & 0.34 & 194 &  2x8 \\
\enddata
\end{deluxetable}


\clearpage

\begin{references}
\reference{}

\reference{} [1] A. R. Thompson, B.G. Clark, C. M. Wade, and P. J. Napier, {\it The 
             Very Large Array}, Astrophys. J. Suppl., vol 44, pp 151-167, 1980. 

\reference{} [2] P. J. Napier, A. R. Thompson and R. D. Ekers, {\it The Very Large 
             Array: Design and Performance of a Modern Synthesis Radio Telescope}, 
             Proc. IEEE, vol. 71, pp 1295-1320, 1983.

\reference{} [3] G. L. James, {\it Design of wide-band compact corrugated horns}, IEEE 
             Trans. Ant. Prop., vol AP-32, pp. 1134-1138, 1984.

\reference{} [4] B. K. Watson, A. W. Rudge, R. Dang and A. D. Olver, {\it Compact low 
             cross-polar corrugated feed for E.C.S.}, IEEE Antennas Propagat. Conf. 
             Digest, Quebec, vol. 1, pp. 209-212, June 1980.

\reference{} [5] S. Srikanth, J. Ruff, and E. Szpindor, {\it Design, prototyping and 
             measurement of EVLA L-band feed horn}, EVLA Memo 87, 2005

\reference{} [6] S. Srikanth and J. Ruff, {\it Design, prototyping and measurement of 
             EVLA S-band feed horn}, EVLA Memo 112, 2007

\reference{} [7] S. Srikanth, J. Ruff, and A. J. Fenn, {\it Design, prototyping and 
             measurement of EVLA C-band feed horn}, EVLA Memo 95, 2005
	
\reference{} [8] B. MacA. Thomas, {\it Design of corrugated conical horns}, IEEE Trans. 
             Ant. Prop., vol AP-26, pp. 367-372, 1978.
 
\reference{} [9] Y. Takeichi, T. Hashimoto, and F. Takeda, {\it The ring-loaded corrugated 
             waveguide}, IEEE Trans. Microwave Theory Tech., vol. MTT-19, pp. 947-950, 
             Dec. 1971.

\reference{} [10] S. J. Skinner and G. L. James, {\it Wide-band orthomode transducers}, 
             IEEE Trans. Microwave Theory Tech., vol MTT-39, pp. 294-300, 1991. 

\reference{} [11] S. Srikanth, {\it A wide-band corrugated rectangular waveguide phase 
             shifter for cryogenically cooled receivers}, IEEE Microwave \& Guided Wave 
             letters, vol.7, pp. 150-152, 1997.  

\reference{} [12] A. M. B{\rm\char'34}ifot, E. Lier and T. Schaug-Pettersen, 
             {\it Simple and broadband orthomode transducer}, IEE Proc., Proc. IEE, 
             vol. 137, no. 6, pp. 396-400, 1990

\reference{} [13] E. J. Wollack, {\it A full waveguide band orthomode junction}, NRAO 
             Electronics Division Internal Report 303, 1996

\reference{} [14] E. J. Wollack, W. Grammer, and J. Kingsley, {\it The B{\char'34}ifot 
             orthomode junction}, ALMA Memo 425, 2002

\reference{} [15] M. H. Chen and G. N. Tsandoulas, {\it A wide-band square-waveguide 
             array polarizer}, IEEE Trans. Ant. Prop., vol AP-21, pp. 389-391, 1973.

\reference{} [16] D. Bock, {\it Measurements of a scale-model ortho-mode transducer},
              BIMA Memo 74, 1999.

\reference{} [17] G. Engargiola and R. Plambeck, {\it Tests of a planar L-band orthomode 
             transducer in circular waveguide}, Review of Scientific Instruments, vol. 
             74, no. 3, March 2003.

\reference{} [18] P. K. Grimes, O. G. King, G. Yassin and M. E. Jones, {\it Compact 
             broadband Orthomode Transducer}, Electronics Letters, vol. 43, no. 21, 
             October 2007.

\reference{} [19] Y. Aramaki, M. Miyazaki and T. Horie, {\it Ultra-thin broadband OMT 
             with turnstile junction}, IEEE MTT-S International Microwave Symposium Digest,
             ISSN 0149-645X, vol. 1, pp. 47-50, 2003.

\reference{} [20] M. W. Pospieszalski, {\it Extremely Low-Noise Amplification with 
             Cryogenic FETs and HFETs: 1970-2004}, Microwave Magazine, vol. 6, no. 3, 
             pp. 62-75, Sept. 2005 

\reference{} [21] J. C. Webber and M. W. Pospieszalski, {\it Microwave instrumentation 
             for radio astronomy}, IEEE Trans. Microwave Theory Tech, vol. MTT-50, no.3, 
             pp 986-995, 2002

\reference{} [22] M. W. Pospieszalski, E. J. Wollack, N. Bailey, D. Thacker, J. Webber, 
             L. D. Nguyen, M. Le and M. Lui, {\it Design and Performance of Wideband, 
             Low-Noise, Millimeter-Wave Amplifiers for Microwave Anisotropy Probe 
             Radiometers}, in Proc. 2000 IEEE MTT-S Int. Microwave Symp., Boston, MA, 
             pp. 25-28, June 11-16, 2000

\reference{} [23] T. Gaier, S. Weinreb, L. Samoska, C. Lawrence, D. Dawson, M. Wells, 
             A. Collins, and D. Price, {\it Amplifier technology for astrophysics},  
             Far-IR, Sub-mm \& MM Detector Technology Workshop, April 2002

\reference{} [24] M. W. Pospieszalski, {\it Modeling of Noise Parameters of MESFETs 
             and MODFETs and Their Frequency and Temperature Dependence}, IEEE Trans. 
             Microwave Theory and Tech., vol. MTT-37, pp. 1340-1350, Sept. 1989.

\reference{} [25] I. Angelov , N. Wadefalk, J. Stenarson, E. L. Kollberg, P. Starski, and 
             H. Zirath, {\it On the Performance of Low-Noise, Low-DC-Power Consumption 
             Cryogenic Amplifiers}, IEEE Trans. Microwave Theory Tech., vol. 50, pp. 
             1480-1486, June 2002.

\reference{} [26] M. Morgan, {\it Millimeter-wave MMICs and Applications}, PhD Thesis, 
             California Institute of Technology, May 2003.

\reference{} [27] M. Morgan and S. Weinreb, {\it Techniques for the integration of 
             high-Q millimeter-wave filters in multi-function modules}, Microwave Journal, 
             May 2005.

\reference{} [28] B. R. Carlson, P.E. Dewdney, T.A. Burgess, R.V. Casorso, W.T. 
             Petrachenko, and W.H. Cannon, {\it The S2 VLBI Correlator: A Correlator 
             for Space VLBI and Geodetic Signal Processing}, Publications of the 
             Astronomical Society of the Pacific, 1999, 111, 1025-1047.

\reference{} [29] B. R. Carlson and P. E. Dewdney, {\it Efficient wideband digital 
             correlation}, Electronics Letters, IEE, Vol. 36 No. 11, p987, 25 May 2000.

\end{references}
\end{document}